\documentclass[sigconf,screen]{acmart}

\usepackage[utf8]{inputenc}
\usepackage{todonotes}
\usepackage{amsmath}
\usepackage{enumerate}
\usepackage{booktabs}
\usepackage{url}
\usepackage{lscape}
\usepackage[font=small,labelfont=bf]{caption}
\usepackage{blindtext}
\usepackage{wrapfig}
\usepackage{multirow}
\usepackage{graphicx}
\usepackage{xcolor}

\AtBeginDocument{%
  \providecommand\BibTeX{{%
    \normalfont B\kern-0.5em{\scshape i\kern-0.25em b}\kern-0.8em\TeX}}}

\setcopyright{none}
\acmConference[RecSys in HR 2021]{Workshop on Recommender Systems for Human Resources at the ACM RecSys Conference on Recommender Systems (RecSys 2021)}{October 1, 2021}{Amsterdam, the Netherlands}
\acmDOI{}
\acmISBN{}
\copyrightyear{2021. Copyright for the individual papers remains with the authors. Copying permitted for private and academic purposes. This volume is published and copyrighted by its editors}
\acmPrice{}
\begin{document}

\title{Job Posting-Enriched Knowledge Graph for Skills-based Matching}

\author{Maurits de Groot}
\authornote{Work done while on internship at Randstad Groep Nederland.}
\email{maurits.degroot@live.nl}
\affiliation{%
  \institution{Leiden University}
  \streetaddress{?}
  \city{Leiden}
  \country{The Netherlands}
}

\author{Jelle Schutte}
\email{jelle.schutte@randstad.com}
\affiliation{%
  \institution{Randstad}
  \streetaddress{Diemermere 25}
  \city{Diemen}
  \country{The Netherlands}
}

\author{David Graus}
\email{david.graus@randstadgroep.nl}
\affiliation{%
  \institution{Randstad Groep Nederland}
  \streetaddress{Diemermere 25}
  \city{Diemen}
  \country{The Netherlands}
}

\renewcommand{\shortauthors}{de Groot, et al.}

\begin{abstract}
The labor market is constantly evolving. 
Occupations are changing, being added, or disappearing to fit the needs of today's market. 
In recent years the pace of this change has accelerated, due to factors such as globalization, digitization, and the shift to working from home. 
Different factors are relevant when selecting employment, e.g., cultural fit, compensation, provided degree of freedom. 
To successfully fulfill an occupation the gap between required (by the job) and possessed (by the job seeker) skills needs to be as small as possible. 
Decreasing this skill-gap improves the fit between a job candidate and occupation. 

In this paper we propose a custom-built Skills \& Occupation Knowledge Graph (KG) that fits the above described dynamic nature of the labor market, by leveraging existing skills and occupation taxonomies enriched with external job posting data. 

We leverage this KG and explore several applications for skills-based matching of jobs to job seekers. 
First, we study link prediction as a means to quantify relevance of skills to occupations, which can help in prioritizing learning and development of employees. 
Next, we study node similarity methods and shortest path algorithms for career pathfinding. 
Finally, we leverage a term weighting method for identifying which skills are most ``distinctive'' for different (types of) occupations. 

\end{abstract}



\maketitle

\section{Introduction}\label{sec:introduction}

In recent years the number of people that change their job is increasing~\cite{Eurostat}, the average duration of a position is shorter~\cite{OECD} and the total working population is growing~\cite{OECD2}. Due to increasing globalization, the number of possible job candidates per position is higher. And candidates enjoy, on average, a higher level of education compared to a number of years ago~\cite{Akcomak2011}. 
This results in a rapidly increasing number of potential job candidates and the labor market is more competitive than it has ever been~\cite{NBERw11094}. 

In addition, with demand of skills changing over time, having the correct skills for specific occupations is more crucial than ever. 
The increasing amount of digitization has made computer skills more valuable~\cite{PENG201726}. 
The COVID-19 pandemic has resulted in a double-disruption effect where technological adoption is accelerated and companies lay off employees~\cite{world2020future}. 
Most aging workers do not posses the newly required technical skills which leads to lower job opportunities~\cite{bosch2013}. 
Not only technical skills are important, having good people skills is becoming increasingly important as well~\cite{doi:10.1177/001979391406700202}. 

The volatility in the labor market results in a change of occupations with new required skills, and being able to keep up with the latest developments is a challenge. 
To find relevant vacancies and job postings, individuals can use external services to match their skills with their desired work. 
In 2019, employment agencies were responsible for fulfilling $10\%$ of the available jobs in the Netherlands ~\cite{cbs1}. 

As explained above, in recent years the labor market has become more competitive, and requirements more dynamic. 
As a result of this, there is a rising interest in skill-based matching of candidates to jobs~\cite{world2020future}, as the desired profiles for a given occupation are no longer static and unambiguous.

\subsection{Problem Statement}\label{sec:PS}

To facilitate candidate to job posting matching, it is important to know which skills are relevant, in demand, and in supply. 
Here, the need for a flexible data representation for skills arises. 
This representation should facilitate various tasks, such as a skills similarity metric to be able to quantify likeliness between skills, skills-to-occupation similarity metrics, to help people navigate the labor market and find new occupations, and understanding which skills relate to which occupations to inform which skills are needed for desired occupations. 
And since relations between skills and occupations are not static and need robust and accurate updating methods to ensure the information does not get outdated.

In this paper we address the task of skills and occupation graph construction which we describe in Section~\ref{chap:datapreperation}, and apply this data representation to the following set of use-cases: 
link prediction for identifying novel skills-occupation relations in Section~\ref{chap:RQ1}, 
skills-based occupational similarity for career pathfinding in Section~\ref{chap:RQ2}, and 
identifying distinctive skills per occupational group for learning \& development in Section~\ref{chap:RQ3}.

\section{Knowledge Graph Construction}\label{chap:datapreperation}

\begin{figure}[h]
\centering
\includegraphics[width=1.0\columnwidth]{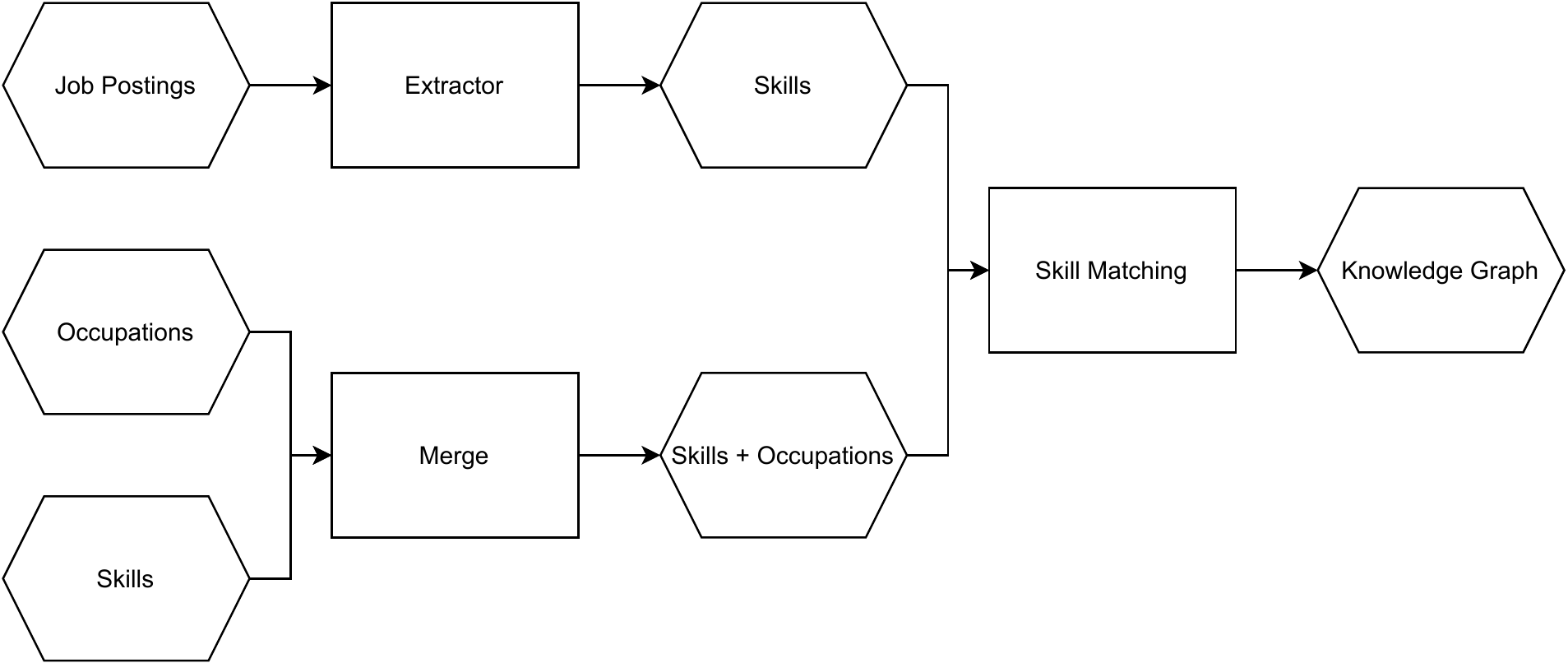}
\caption{Knowledge Graph creation flow}
\label{fig:flow}
\end{figure}

Our Skills \& Occupational KG is based on existing structured data, more specifically, we combine the ISCO (occupations) and ESCO (skills) taxonomies (bottom row in Figure~\ref{fig:flow}).
Next, we enrich this existing data with information from noisy, unstructured job postings (top row in Figure~\ref{fig:flow}) to ensure our KG represents the current state of the labor market.

\subsection{Occupations (ISCO) and skills (ESCO)}
The first step involves constructing a shared Skills \& Occupational Knowledge Graph, through combining the existing ISCO and ESCO taxonomies. 

\subsubsection{ISCO (occupations)}
The International Standard Classification of Occupations (ISCO) is ordered as a taxonomy of occupational groups with four granularity levels across ten different major groups. 
An occupation is defined as \textit{``a set of jobs whose main tasks and duties are characterized by a high degree of similarity''}, where a job is defined as \textit{``a set of tasks and duties performed, or meant to be performed, by one person, including for an employer or in self-employment.''~\cite{ilo}}
Take, for example: the occupation ``computer programmer,'' which is defined by the level 4 ISCO code: 2132. 
The occupation then belongs to the the level 3 group ``computing professionals'' (ISCO-code 213), which in turn belongs the level 2 group ``computing, engineering and science professionals'' (ISCO-code 21),
which, finally, falls in the level 1 group ``professionals'' (ISCO-code 2). 

\begin{table}[h]
\centering
\begin{tabular}{@{}ll@{}}
\toprule
Group Number & Major Group Name                                   \\ \midrule
1            & Managers                                           \\
2            & Professional                                       \\
3            & Technicians and associate professionals            \\
4            & Clerical support workers                           \\
5            & Service and sales workers                          \\
6            & Skilled agricultural, forestry and fishery workers \\
7            & Craft and related trades workers                   \\
8            & Plant and machine operators, and assemblers        \\
9            & Elementary occupations                             \\
10           & Armed forces occupations                           \\ \bottomrule
\end{tabular}
\caption{The 10 major job groups of the ISCO-08}
\label{tab:isco}
\end{table}

\subsubsection{ESCO (skills)}
We define our initial high-level occupation groups by using the ISCO standard.
For skills, we turn to The European Skills, Competences, Qualifications and Occupations (ESCO) taxonomy~\cite{ESCO}. 
ESCO defines a skill as follows:

\begin{description}
\item[Skill] \textit{``the ability to apply knowledge and use know-how to complete tasks and solve problems''}
\end{description}

The ESCO covers 13,485 skills, connected to 2,942 occupations (in 27 languages). 

We link our ISCO occupations to ESCO by using the direct links that are defined between ISCO level 4 groups (most fine-grained/lowest level of the taxonomy) and ESCO concepts, in the ESCO. 
These links between ESCO and ISCO are not (necessarily) 1-to-1, as multiple ESCO occupations can be linked to a single (level 4) ISCO group. 

\begin{figure}[t]
\centering
\includegraphics[width=1.0\columnwidth]{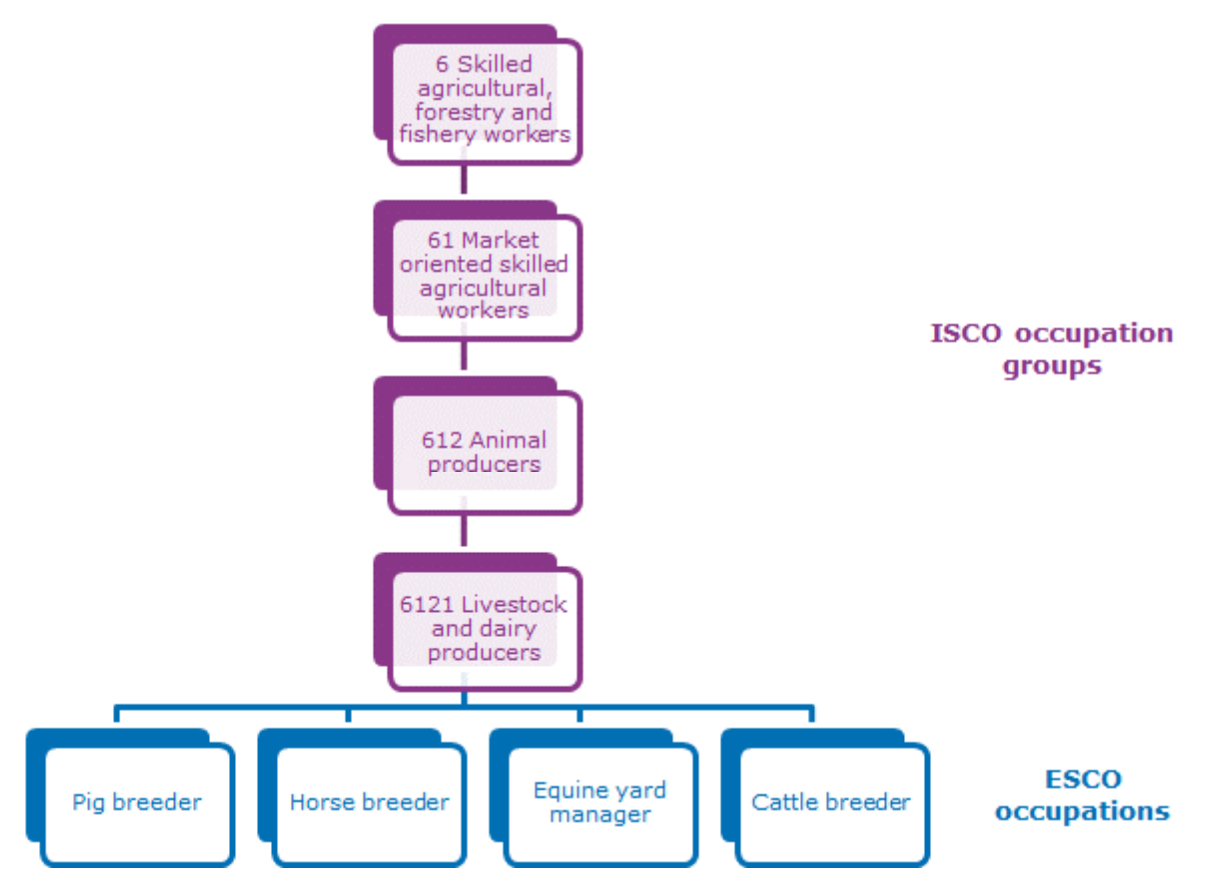}
\caption{The structure of the occupations pillar \cite{ESCO}}
\label{fig:TheStructureOfTheOccupationsPillar}
\end{figure}

In Figure~\ref{fig:TheStructureOfTheOccupationsPillar} we illustrate this connection between ISCO and ESCO. 
ESCO occupations are shown in blue, with ISCO occupation groups in purple. 
In addition to the ESCO occupations shown in the image, ESCO also defines skills (not shown), e.g., the ESCO occupation ``Cattle breeder,'' has skills linked to them such as ``feed livestock'' and ``assist animal birth.'' 

\subsection{KG enrichment through job posting data}
Now that we have our high-level KG structure based on ISCO and ESCO, which defines occupations and skills as nodes, and edges as links between ESCO and ISCO objects, we turn to job posting data to account for the dynamic nature of associations between skills and occupations, as described in Section~\ref{sec:introduction}. 
To make sure our KG reflects the current status of the labor market, we use information from job postings to enrich the structure of our KG. 
More specifically, we create additional edges by identifying and extracting ESCO skills for each job posting's ISCO occupation group, and assign weights to edges by relying on co-occurrence statistics of skills and occupations. 

This second step of our process revolves around extracting skills from job postings. 
We describe our job posting dataset in Section \ref{Vacancy Parser}, our approach for skill extraction in Section \ref{Vacancy Parser}, and how we match extracted skills to ESCO skills in~\ref{Skill Matching}.

\subsubsection{Vacancy data}\label{Vacancy data}
Our vacancy dataset consists of sample of 600,000 Dutch vacancies collected by Jobdigger~\cite{Jobdigger}, each job posting is labeled with a level 4 ISCO code. 
Our sample was chosen by selecting a uniform distribution of ISCO level 1 occupations, to make sure our set covers the entire breadth of the labor market. 
Prior to sampling our set at the ISCO level 1, the initial dataset was cleaned by discarding low quality and noisy job postings, such as postings that represented multiple occupations, or job postings that contained a low number of sentences. Here, we treat vacancy data as a proxy for the demand in the job market. By doing so, internal promotions and career paths and informal channels are not taken into account. 

\subsubsection{Skill Extraction}\label{Vacancy Parser}
For skill extraction we rely on the %
industry-standard Textkernel Extract~\cite{TextKernel} parser. 
For each vacancy text, Textkernel Extract returns a json object with corresponding skills, represented by the surface form identified in the job posting (skill mention), a unique identifier representing the skill (skill id), and finally, a confidence score that quantifies the likelihood of the extracted skill to be correct.

\subsubsection{Skill Matching}\label{Skill Matching}

\begin{figure}[t]
\includegraphics[width = 0.8\columnwidth]{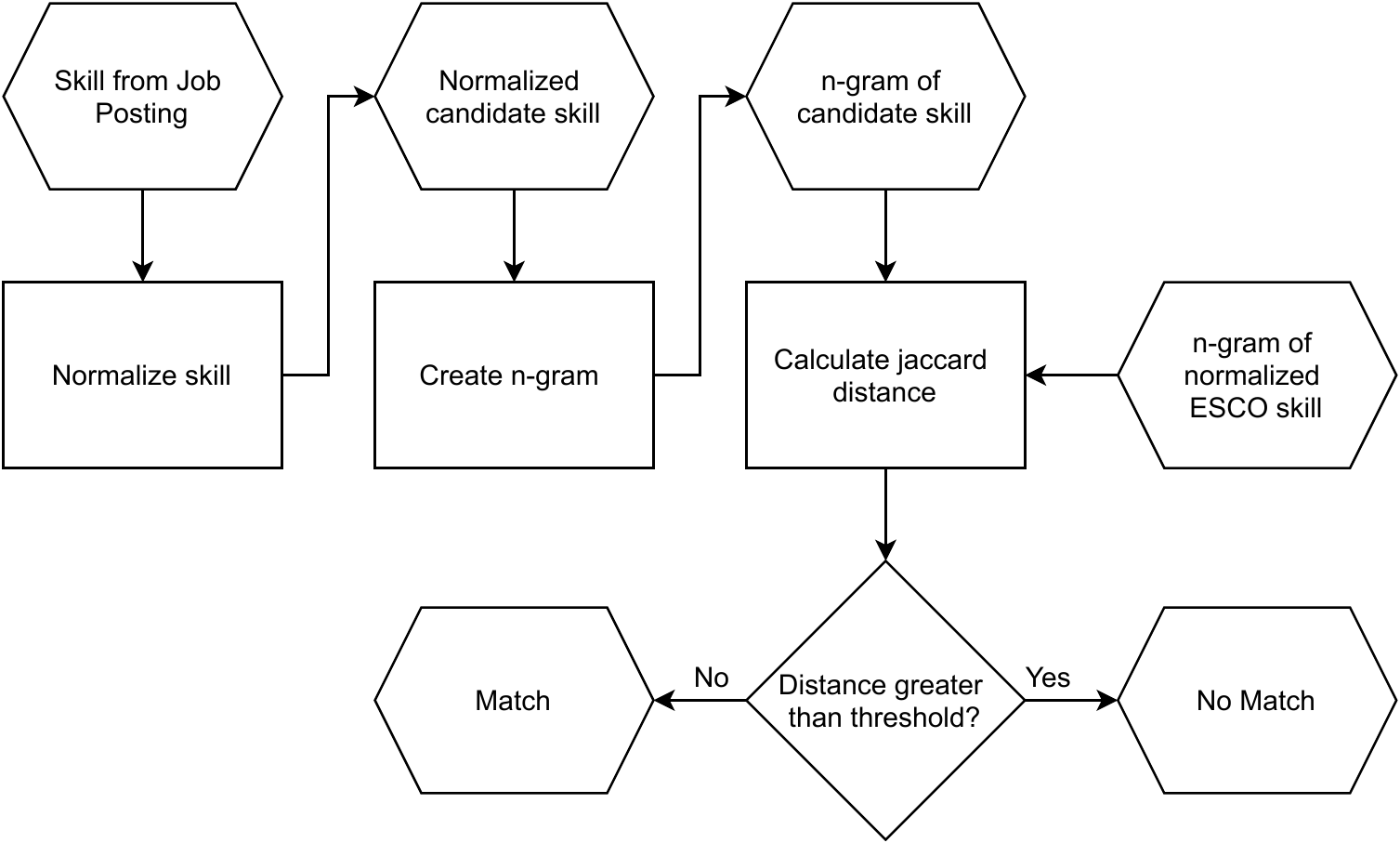}
\caption[Overview of skill matching process]{Overview of skill matching process}
\label{fig:skill_match_flow}
\end{figure}

Given the skills extracted by Textkernel, we match them to the skill nodes in our KG, by relying on the surface forms of the skills (skill mentions).
More specifically, we leverage character $n$-grams Jaccard similarity between the normalized skill mention and the normalized ESCO skill names. 
We set the similarity threshold to $0.66$, which was empirically determined to be optimal using a smaller set of our $39,758,827$ Textkernel skills to ESCO skill-mappings. 
The high-level process is shown in Figure~\ref{fig:skill_match_flow}. 

\subsection{Final Skills \& Occupational Knowledge Graph}
Our final KG, resulting from the process shown in Figure~\ref{fig:flow} and described in the previous section, consists of 1,220 nodes, of which 983 represent (ESCO) skills, and 237 (ISCO) occupations. 
These nodes are connected through $3,910$ edges, with an average node degree of $6.4$. 

This KG is a subset of the full ESCO ($13.485$ skills), and ISCO ($436$ occupations) taxonomies. 
There are several reasons why our KG is a subset and does not span the entirety of the ISCO and ESCO taxonomies.

First, it is conceivable that not all ISCO occupations are in current demand, e.g., we found that there were no vacancies for ISCO occupation code 8111: ``mining-plant operators,'' which is not surprising with currently no mines in operation in The Netherlands. 
Next, it is likely we are dealing with coverage issues, from
(i) the likely incomplete coverage of the TextKernel Extract method we use for skill extraction, and 
(ii) our skills matching methodology further reducing the number of identified skills. 
As the focus of this paper is on downstream applications, we consider matching out of scope, and rely on our naive but solid character $n$-grams-based method.

\section{KG Completion using Link Prediction}\label{chap:RQ1}
One of the challenges of modeling skills and occupations is the dynamic nature of the labor market. 
In this section we explore our first down-stream application of our data-driven dynamically constructed Skills \& Occupation Knowledge Graph: matching occupations to skills. 
We focus on discovering novel connections between skills and occupations through leveraging the structure of our knowledge graph enriched with job posting data. 

More specifically, in this section we compare link prediction algorithms, to quantify the relatedness between a skill and occupation node, in order to discover novel connections between skills and occupations, not present in our initial KG. 
We describe our two link prediction methods in the following sections, 
the first, Preferential Attachment, is described in Section~\ref{sec:method_pa}, next, Node2Vec is described in Section~\ref{sec:method_n2v}

\subsection{Experimental setup}
We employ link prediction to estimate the relatedness between skills and occupation nodes. 
To evaluate and reliably compare different methods, we first split our KG into train, test, and validation sets. 
More specifically, we sample 55\% of all edges for training the link prediction algorithms (where applicable), leaving leave 30\% for testing, and 15\% for validation. 
For each existing pair of occupation and skills node --- which we consider a positive sample in our train, test and validation sets --- we randomly generate a negative sample (i.e., a pair of skills and occupation nodes that do not exist in our KG). 
An overview of the number of edges in each set is shown in Table~\ref{tab:train_val_test}.

\begin{table}[!h]
\centering
\begin{tabular}{@{}lrr@{}}
\toprule
                 & Positive & Negative \\ \midrule
Training edges   & 2151     & 2151     \\
Validation edges & 586      & 586      \\
Test edges       & 1173     & 1173     \\ \midrule
Total            & 3910     & 3910     \\ \bottomrule
\end{tabular}
\caption[Number of positive and negative edges]{Number of positive and negative edges with a training (55\%), validation (15\%), test (30\%) split }
\label{tab:train_val_test}
\end{table}

\subsection{Link Prediction Methods}
\subsubsection{Method 1: Preferential Attachment (PA)}
\label{sec:method_pa}
The first link prediction method is preferential attachment~\cite{liben2007link}. 
This method takes a set of nodes, i.e. node $v$ and node $u$, and calculates a closeness ($C$) between two nodes:

\begin{equation}
C(u,v) = |\Gamma(u)| \times |\Gamma(v)|,
\end{equation}
where $\Gamma(u)$ denotes the neighbors of $u$.

A higher score here corresponds to a larger probability the nodes are connected. 
The intuition behind this is that if both nodes have a high amount of neighbors the nodes might function as a hub. 
Most graphs have the property that hubs have a higher chance to be connected. 

To compute all scores, we represent our KG as a matrix, where each node is represented as a row and a column.
Note that this matrix is symmetric since the value for row $u$ and column $v$ is equal to the value at row $v$ and column $u$. 
At the intersecting cell of two nodes, we store the preferential attachment. 
We normalize this matrix by dividing each score by the maximum Closeness score, to ensure that each value is between 0 and 1. 
We consider the resulting normalized Closeness score as the probability the corresponding nodes are related. 

\subsubsection{Method 2: Node2Vec (N2V)}
\label{sec:method_n2v}
The second link prediction method we use is the Node2Vec algorithm~\cite{grover2016node2vec}. 
This algorithm can have a number of configurations. 
For this paper we use the following parameters:

\begin{itemize}
\item dimensions = 1024
\item walk length = 4
\item number of walks = 2500
\item $p$ (return parameter) = 1 
\item $q$ (in-out parameter) = 1
\end{itemize}

These parameters were selected after a grid search on a large number of possible combinations of parameters. 

\subsection{Results}
Table \ref{tab:F1-results} shows the performance of both Preferential Attachment (PA) and Node2Vec (N2V).

\begin{table}[!h]
\centering
\begin{tabular}{@{}llllll@{}}
\toprule
                                         & class    & precision & recall & f1-score \\ \midrule
\multirow{2}{*}{PA} & 0.0      & \textbf{0.83}      & 0.64   & 0.72    \\
                                         & 1.0      & 0.71      & \textbf{0.87}   & \textbf{0.78}    \\ \midrule
\multirow{2}{*}{N2V}                & 0.0      & 0.66      & \textbf{0.90}   & \textbf{0.76}    \\
                                         & 1.0      & \textbf{0.84}      & 0.53   & 0.65    \\ \bottomrule
\end{tabular}
\caption[Precision, recall and F1-scores of multiple link prediction algorithms]{Precision, recall and F1-scores of multiple link prediction algorithms with an equal number of positive and negative edges used for training}
\label{tab:F1-results}
\end{table}

\begin{figure}[!h]
\centering
\includegraphics[width=\columnwidth]{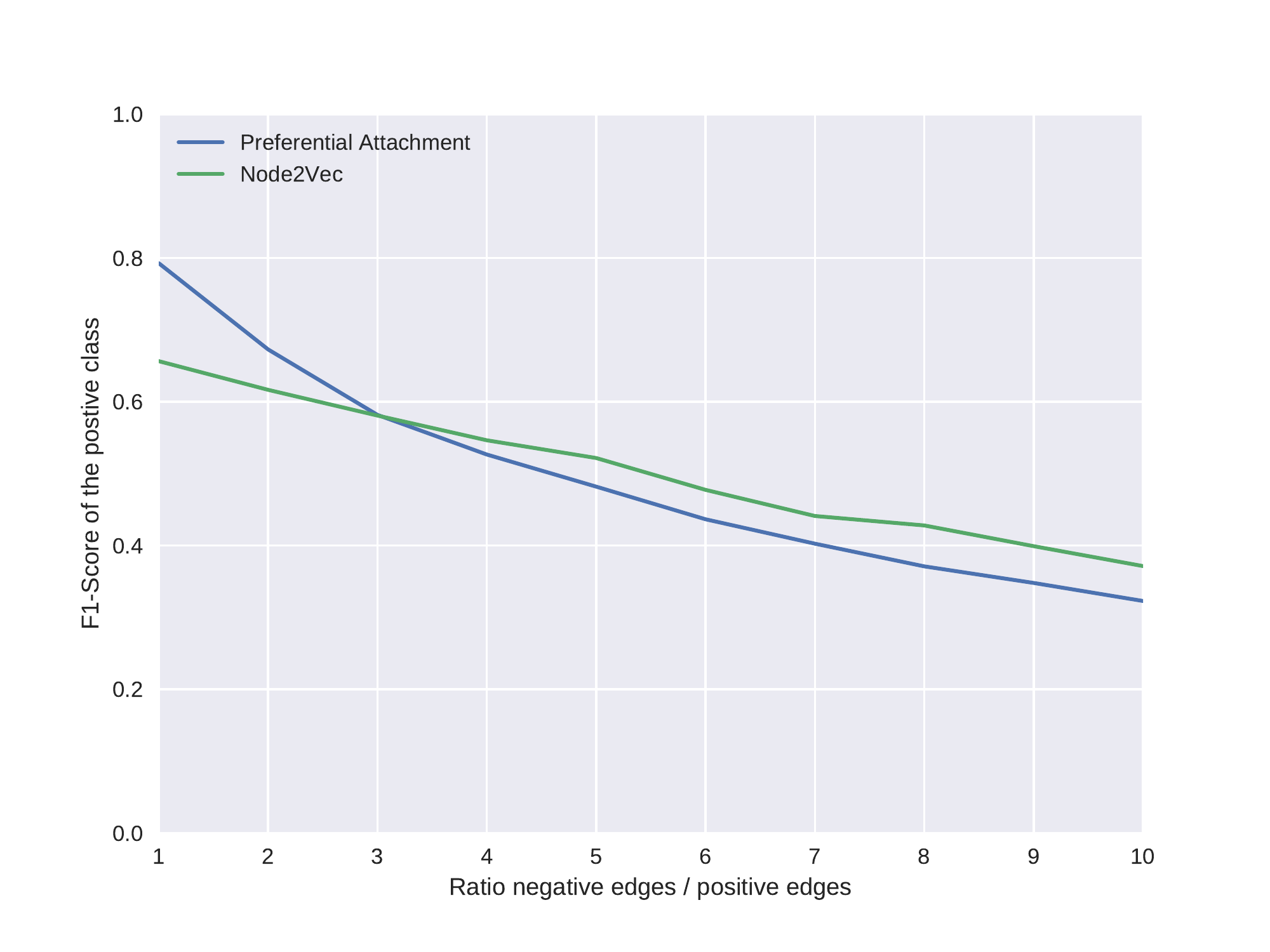}
\caption[Comparison of Node2Vec and Preferential Attachment]{Comparison of Node2Vec and Preferential Attachment for different ratio's negative edges / positive edges}
\label{fig:f1-score}
\end{figure}

When the number of positive and negative edges in the test set is equal, PA outperforms the more complex N2V method, with an f1-score for the positive class of $0.78$ against $0.65$. 
In most realistic situations however, we may want to explore how a node can be linked to any other node, making the number of comparisons, or edges to predict 1-to-(N-1), i.e., for each node we compare each other node (excluding self). 
To approximate this real world performance the ratio of negative to positive edges should reflect these more realistic proportions. 
To do so we compute F1-score at increasing ratios of positive-to-negative edges, ranging from 1 (as shown in Table~\ref{tab:F1-results}) to 7. 
Results are shown in Figure~\ref{fig:f1-score}.
The figure shows that up to ratio of 3:1, N2V is on par with PA, but as ratios increase, N2V outperforms PA, suggesting N2V is better suited for most real world situations.

\subsection{Analysis}
Now that it has been established that N2V is more suitable for our task, we aim to employ this algorithm to predict the relationships between occupations and skills. 
When doing so we need to realize that the graph which we use as input is imperfect in terms of correctness and completeness~\cite{paulheim2017knowledge}. 

Looking at the false positives of the algorithm, skills that are --- according to our dataset --- incorrectly linked to occupations can be identified. 
For KG completion, we aim to identify those skills that are not linked to occupations, but should be. 
Table~\ref{tab:example_node2vec} shows a random sample of False Positives: it reinforces our intuition that link prediction can be employed for KG completion, as some of the predicted edges make sense, e.g., the skill: ``preparing materials for dental procedures'' is shown as a relevant skill for the occupation: ``dentist.''
By consulting domain experts, skills can be efficiently added to enrich the current graph. 

\begin{table*}[t]
\centering
\begin{tabular}{@{}lll@{}}
\toprule
ISCO-Code & Occupation & Predicted Skill \\ \midrule
1341  & Child care services managers  & children's physical development \\
2261  & Dentists  & prepare materials for dental procedures \\
3251  & Dental assistants and therapists  & dentistry science \\
4110  & General office clerks  & demonstrate professional attitude to clients \\
5411  & Fire fighters  & safety engineering \\
6121  & Livestock and dairy producers  & promote animal welfare \\
7132  & Spray painters and varnishers  & spray pesticides \\
8344  & Lifting truck operators  & hazardous materials transportation \\
9111  & Domestic cleaners and helpers  & provide lawn care \\ \bottomrule
\end{tabular}
\caption{False positives: edges predicted by N2V that do not exist in our KG}
\label{tab:example_node2vec}
\end{table*}

\begin{figure}[h]
\centering
\includegraphics[width = \columnwidth]{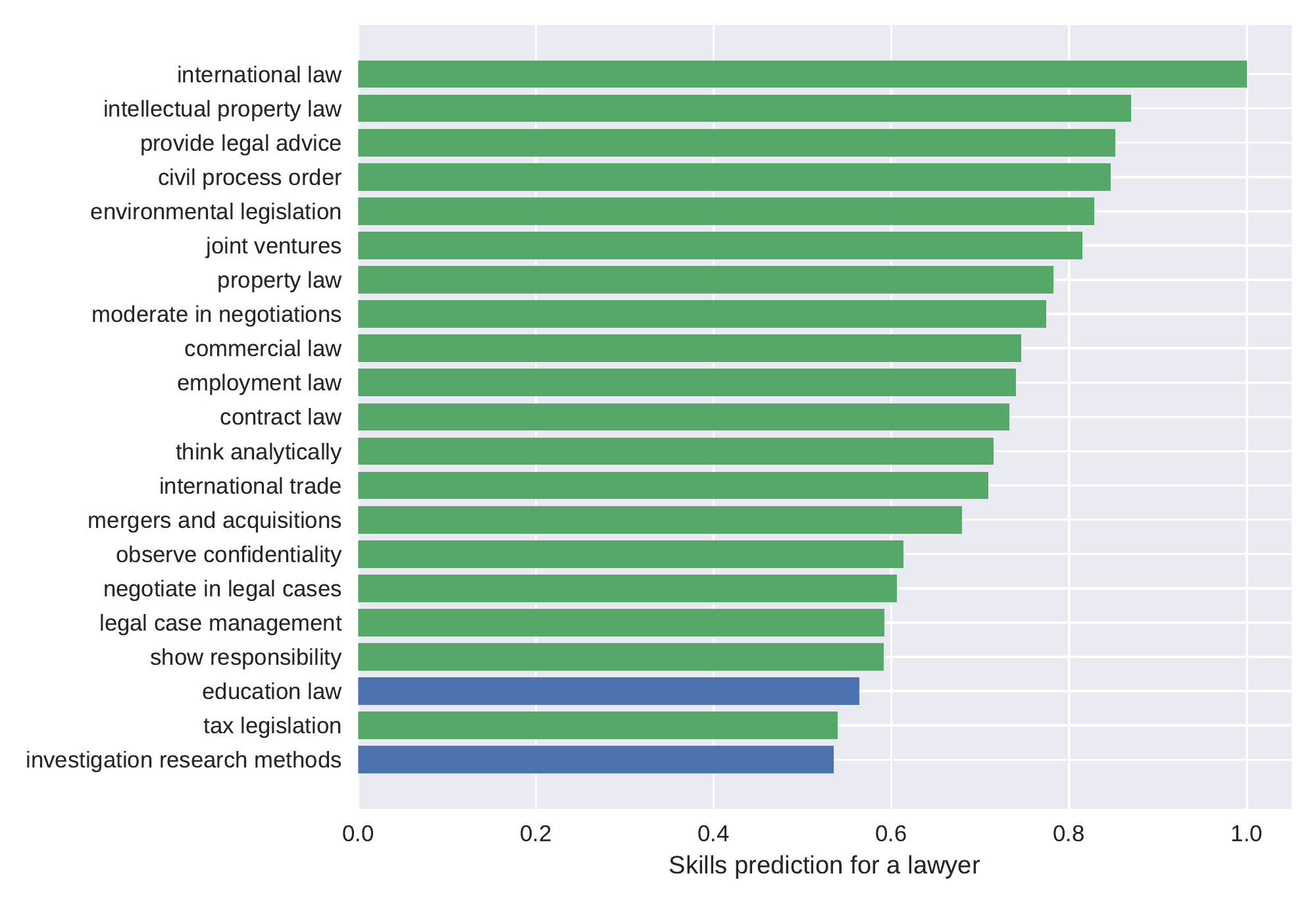}
\caption[Predictions of the Node2Vec algorithm]{Predictions of the Node2Vec algorithm for ISCO group $2611$ (Lawyers)}
\label{fig:tb}
\end{figure}

To further explore these intuitions, in Figure~\ref{fig:tb} we show the edges to skill nodes predicted by N2V, for the node representing ISCO code 2611: ``Lawyers.'' 
The y-axis shows skills edges, and the x-axes show the link prediction probabilities, for all predictions with a probability$>$0.5 (i.e., positive predictions by the method). 
The green bars denote True Positives (i.e., correctly predicted edges between the skill and occupation), and blue bars depict False Positives (skills that are predicted to have an edge with the occupation, but do not exist in our KG). %
The figure shows ``education law'' and ``investigation research methods'' as newly identified skills for lawyers, not found in the original ESCO taxonomy nor in co-occurrences in job postings. 

\section{Career Pathfinding using Shortest Path Algorithms}\label{chap:RQ2}
According to recent data (2019) 1.1 million people switched occupation in the Netherlands \cite{cbs1}. 
When transitioning between one job to another, the gap between both jobs cannot be too large. 
This gap can be considered too large if the required skills for one, differs too much from the other. 
Consequently, occupations that share a large number of skills should be easier to transfer between. 
In this chapter we focus on leveraging skills for better informing transitions between occupations.
More specifically, we aim to leverage the KG structure for matching occupations with occupations, to identify how an individual can change jobs in the most optimal way. 

\subsection{Skills-based Occupation Similarity}
\begin{figure}[h]
\centering
\includegraphics[width = 0.6\columnwidth]{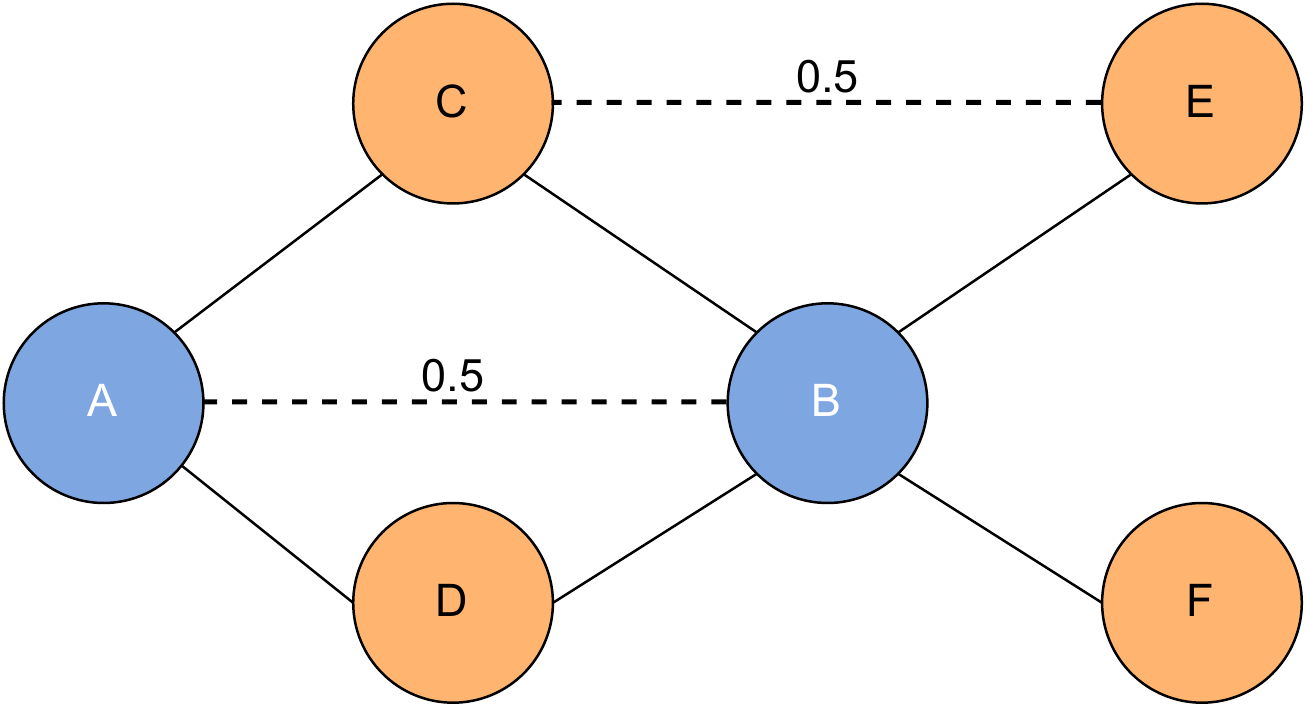}
\caption[Jaccard distance in a graph]{Jaccard distance in a graph where nodes \{A, B\} are occupations and nodes \{C, D, E, F\} are skills. Solid lines denote direct connections, dashed lines denote Jaccard distance.}
\label{fig:jaccard}
\end{figure}

To determine the feasibility of an occupation transfer, we propose to model the distance between occupations with Jaccard distance. 
We compute Jaccard distances between occupations by representing each occupation as the set of its required skills (which we extract from our KG), and computing the overlap between two sets of skills. 
See Figure for an illustration~\ref{fig:jaccard}.

\begin{figure}[!h]
\centering
\includegraphics[width = \columnwidth]{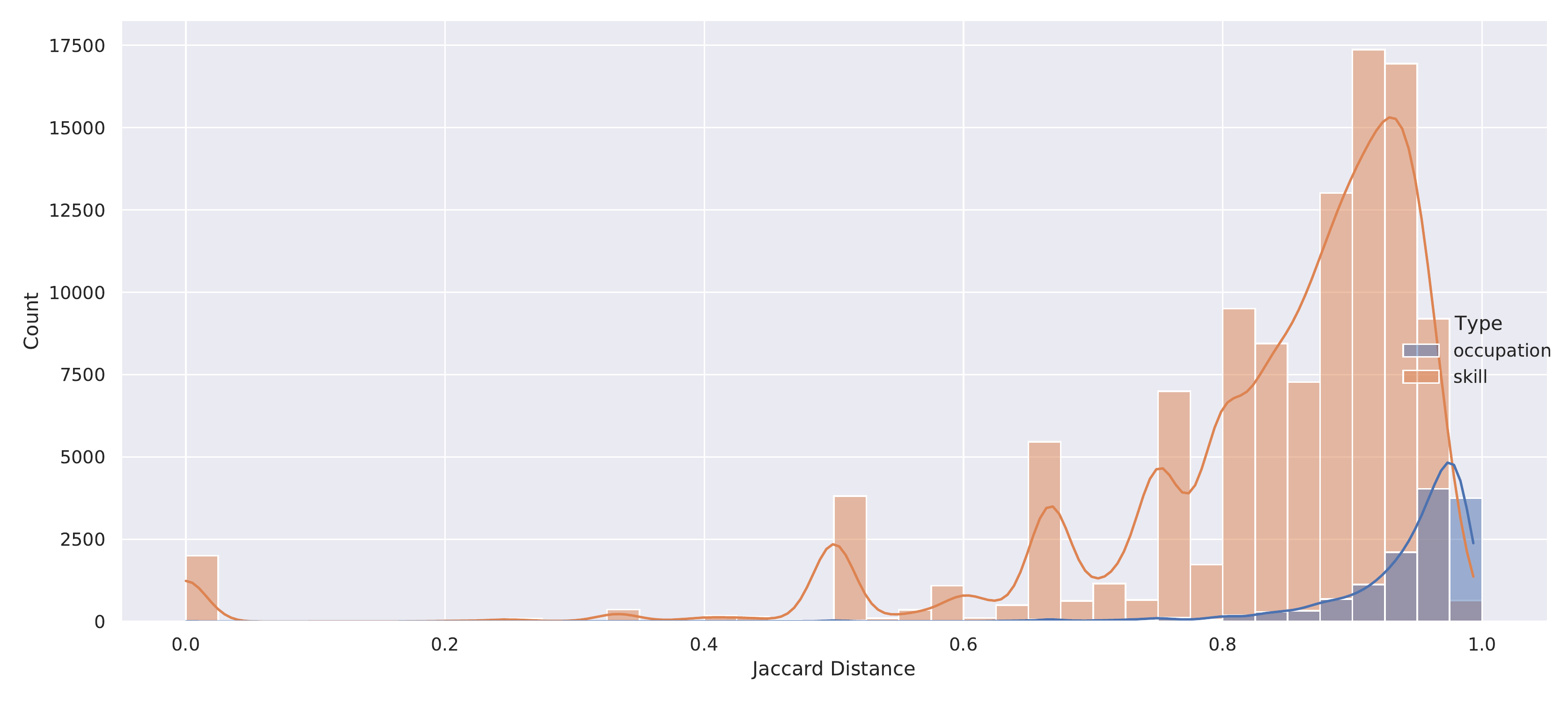}
\caption[Distribution of the jaccard distance ]{Distribution of the jaccard distance where the orange color represents the skills and the blue color represent the occupations}
\label{fig:jaccard_dist}
\end{figure}

In our KG a total of $120,952$ links can be made between pairs of skills and pairs of occupations. 
From these pairs $89.3\%$ is between skills and $10.7\%$ between occupations. 
To gain insight in the overall similarity of skills and occupations, we study the distribution of jaccard distances in Figure~\ref{fig:jaccard_dist}.

Looking at the distribution of Jaccard distance one can see that on average, skills are more similar to one another than occupations. 
This becomes apparent when looking at the mean value of both distributions: for occupations the mean is $0.96$, and for skills around $0.88$. 
Over 99\% of occupations have a Jaccard distance between 0.8 and 1, meaning that occupations require distinct skillsets. 
Both distributions are skewed to the left, meaning that the mean (average of the observations) is left of the mode (most observed value). 

In the distribution we see a number of spikes, which can be explained by the prevalence of some fractions over others, e.g., if half of the neighbors are shared, the Jaccard distance will be $\frac{1}{2}$, which can be achieved in a number of different ways. 
Other spikes occur at additional common fractions such as $\frac{2}{3}$ and $\frac{3}{4}$.

\begin{table}[!h]
\begin{tabular}{@{}lrrr@{}}
\toprule
      & Skill  & Occupation & Total  \\ \midrule
count & 107959 & 12993      & 120952 \\
mean  & 0.825  & 0.938      & 0.837  \\
std   & 0.163  & 0.070      & 0.160  \\
min   & 0.000  & 0.000      & 0.000  \\
25\%  & 0.800  & 0.928      & 0.800  \\
50\%  & 0.875  & 0.960      & 0.888  \\
75\%  & 0.923  & 0.977      & 0.933  \\
max   & 0.985  & 0.993      & 0.993  \\ \bottomrule
\end{tabular}
\caption{Statistics of the jaccard distribution}
\label{tab:jacard_distibution}
\end{table}

In Table \ref{tab:jacard_distibution} we show a description of the distance distributions. 
For both skills and occupations the minimum distance is 0, meaning that a skill is shared by every occupation where the skill is connected to or that two occupations share every skill. 
An example is \textit{``Food service counter attendants''} and \textit{``Hotel receptionists,''} both share the same skillset and thus have a Jaccard distance of 0. 
Skills with a distance of 0 are for example \textit{``Lop trees''} and \textit{``Pruning techniques.''} 
The highest distance found in the dataset is $0.993$, this corresponds with the occupations \textit{``Electronics engineers''} and \textit{``Policy administration professionals.''} 
They share at least one skill but are --- next to the shared skill --- completely different. 
The common skill in this example is \textit{``perform project management.''} 

\subsection{Career Pathfinding using Dijkstra's algorithm}
\begin{figure}[h]
\centering
\includegraphics[width=0.5 \columnwidth]{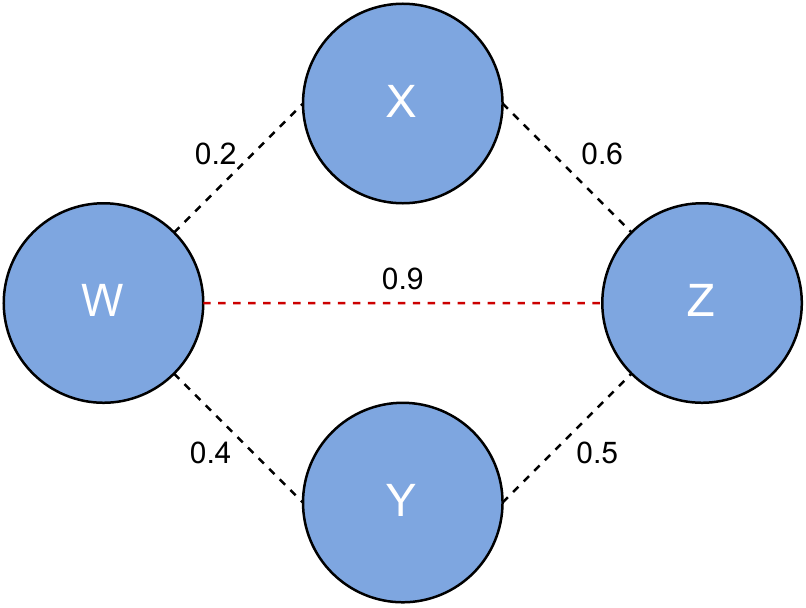}
\caption[Example of distance between occupations]{Distance between the occupations \{$W,X, Y, Z$\}. Black lines denote distances lower than $0.8$. Red lines denote distances higher than $0.8$.}
\label{fig:path_example}
\end{figure}

With the distances between each occupation and between skills, we can proceed to identify the most efficient transition between every pair of occupations. 
This is done by assigning the Jaccard distance scores as edge weights between nodes in our graph, to enable computational methods for finding the most efficient path between a start node (the current occupation) and an end node (the desired occupation). 
We show an example of such a transition in Figure~\ref{fig:path_example}: here we set a threshold for the maximum possible distance at $0.8$. This threshold was determined to be optimal based on eye-balling and comparing a different cutoff points.
If two occupations are further apart than $0.8$ we consider the step too large. 

In this example we start at node $W$ and want to go to node $Z$. 
We are not able to directly transition between $W$ and $Z$ because the occupations are not similar enough ($0.9>0.8$). 

\subsubsection{Method}
Finding the most efficient path in an undirected weighted graph can be done by applying shortest path algorithms. 
For this paper we turn to Dijkstra's algorithm~\cite{dijkstra1959note}, because of its proven speed and widespread availability of implementations. 
According to Dijkstra's algorithm, the shortest allowed path between $W$ and $Z$ in Figure~\ref{fig:path_example} is via node $X$. 

\begin{figure}[h]
\centering
\includegraphics[width=1\columnwidth]{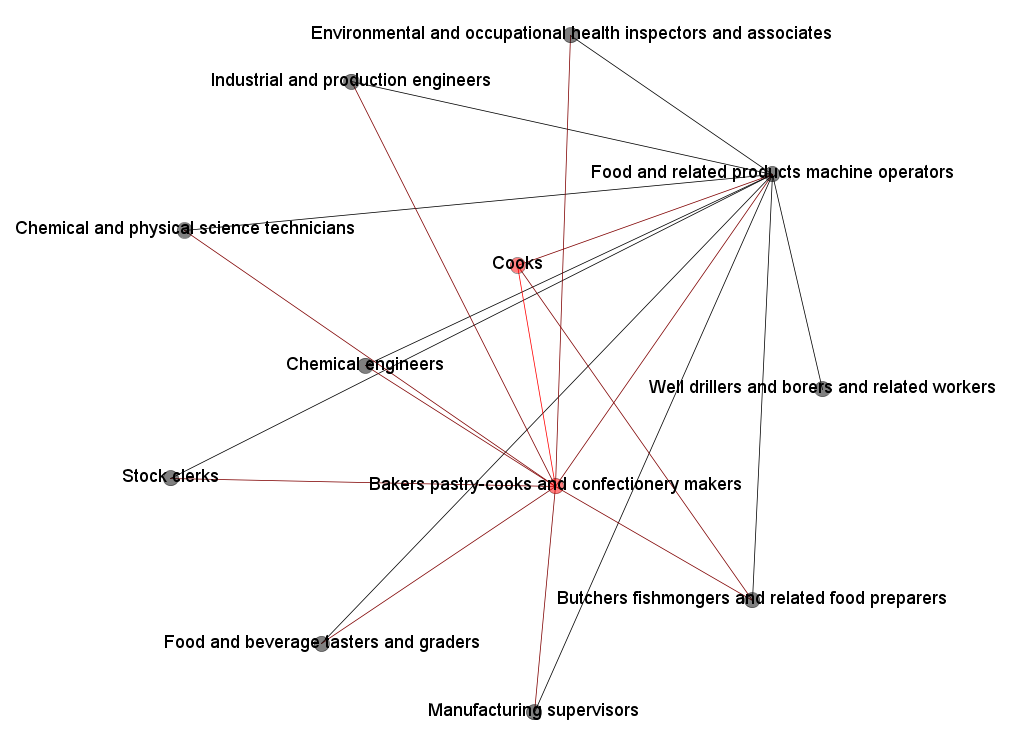}
\caption[The shortest path between occupations]{The shortest path between the occupation ``Cook'' and the closest connected occupation, in this case ``Bakers, pastry-cooks and confectionery makers.''}
\label{fig:covid_example}
\end{figure}

We show a real world example in Figure~\ref{fig:covid_example}. 
Due to the COVID-19 pandemic a lot of people find themselves out of a job, especially individuals that work in restaurants. 
Using the described model we can calculate which occupation has the smallest distance to the occupation: ``cook.'' 
Dijkstra's algorithm yields ``bakers, pastry-cooks and confectionery makers'' as most feasible transition. 

\section{Most Relevant Skills per Occupation Group}\label{chap:RQ3}
Next to fine-grained analysis of occupations and skills, gaining macro-level insights is an important task for monitoring and understanding the labor market. 
The ISCO taxonomy provides multiple levels of granularity, which allows us to aggregate the information contained in our KG at different levels, too. 
In this section we explore a method for identifying the most relevant skills occupations (ISCO level 4) and aggregation of occupations (ISCO level 1-3). 
More specifically, we match skills to occupations at an aggregated level. 

As we've seen in the previous section, different occupations may share skills. 
Several skills, such as \textit{teamwork}, are commonly required for a large number of occupations, which can be considered \emph{generic} or sector-independent skills. 
At the other side, we may have highly specialized skills, that are only required for specific occupations or occupation groups. 
Whether a skill is specifically or generically important can be quantified in different ways. 
For a skill to be specific to an occupation or occupational group, we define two criteria:

\begin{itemize}
\item A skill needs to be 
\emph{frequently required} within its context (occupation or occupation group).
\item A skill needs to be \emph{characteristic} for its context. 
\end{itemize}

\subsection{Method}
The two criteria described above fit naturally to the Term Frequency–Inverse Document Frequency (TF-IDF) weighting scheme for terms~\cite{schutze2008introduction}. 
This statistic is chosen as it directly models the desired criteria described in the previous section, more specifically, TF-IDF is used to assign weights to words in a corpus of documents, where a word is deemed more important if it (i) is observed frequently within the document but (ii) not frequently across different documents in the corpus. 

\begin{equation}
\displaystyle TF-IDF(t,d) = tf_{t,d} \times \log\Big(\frac{N}{df_{t} + 1}\Big),
\end{equation}
where $tf_{t,d}$ denotes the Term Frequency of $t$ in $d$, $df_t$ denotes the number of documents containing $t$, and $N$ denotes the total number of documents in the corpus.

We ``transplant'' this TF-IDF weighting scheme from terms in documents to skills associated to occupations. 
TF-IDF consists of two parts: 
Term Frequency (TF) is the frequency of a word (skill) used in a given document (observed with an occupation), 
Inverse Document Frequency (IDF) is a way to discount highly common terms, i.e., it is high when a word (skill) appears in a smaller number of documents (observed with a low number of occupations). 
Common terms (skills) will thus yield a lower IDF score. 

For our TF-IDF-based model, we consider skills identified in job postings terms, and documents can be modeled as a collection of job postings belonging to an ISCO group. 
The counts of skills, which model term frequency, correspond to the number of times a skill is found in a job posting associated to a certain ISCO code.

\subsection{Results and analysis}
\subsubsection{Level 1 ISCO groups}
The resulting score provides us with skills that are common for a given occupation (group) but uncommon in all other occupation(s) (groups). 
Table~\ref{tab:skill_relevance} shows the top 5 skills for the level 1 ISCO groups. 

In this table \textit{Microsoft Office} appears both in the \textit{Managers} and \textit{Clerical support workers} groups.
For this skill to score high in multiple contexts (occupation groups) the frequencies need to be substantial in both, to be able to compensate for the IDF component of the metric. 
In the Managers group, \textit{Microsoft Office} has a TF of 9\% and in \textit{Clerical support workers} a TF of 5\%. 

\subsubsection{Multiple ISCO levels}
\begin{figure}
\centering
\includegraphics[width=\columnwidth]{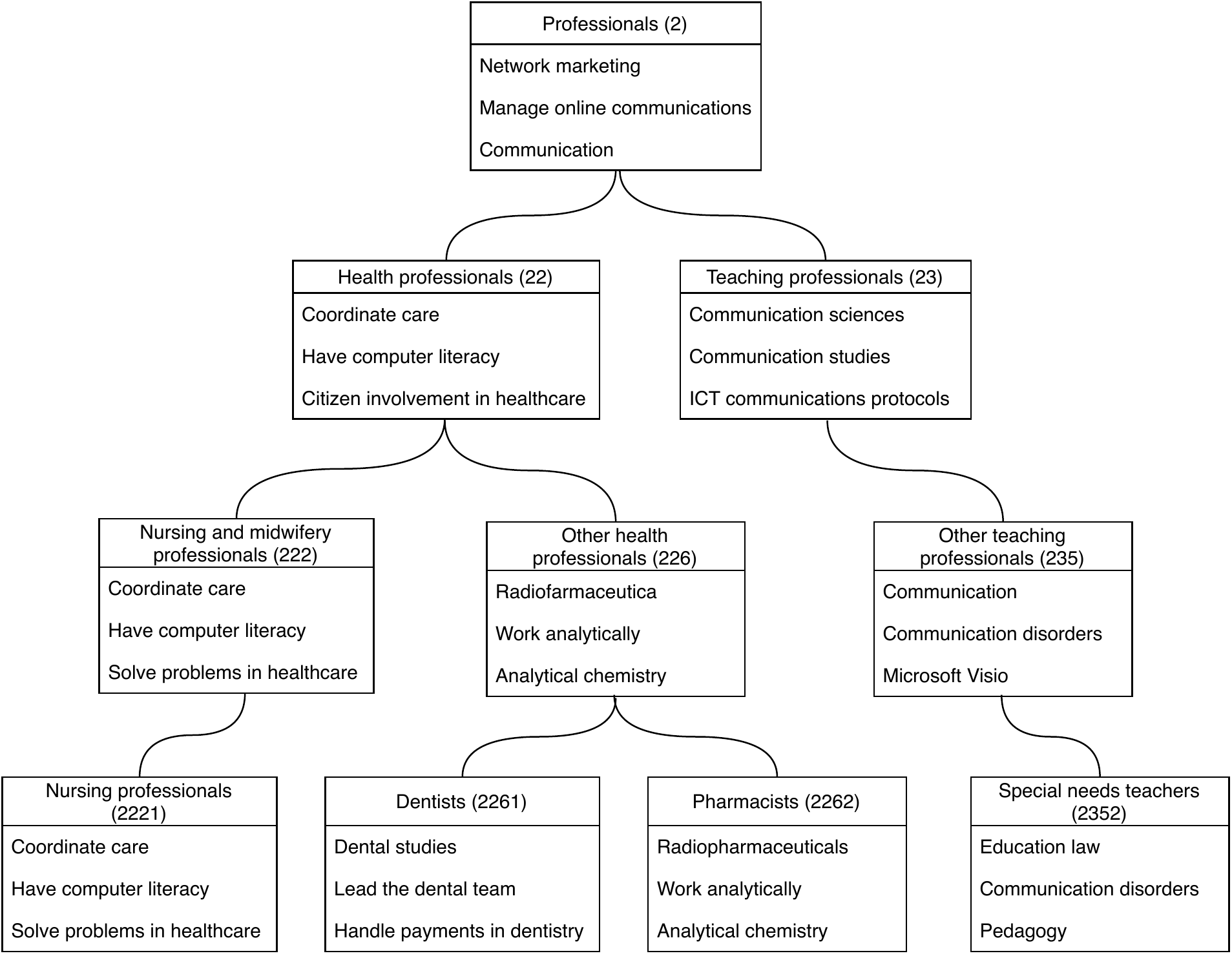}
\caption{Three most relevant skills for multiple levels in major ISCO group 2}
\label{fig:skill_relevance}
\end{figure}

ISCO level 1 helps us to understand which skills are relevant for the least granular level; to deepen our understanding we look at the development of multiple layers of ISCO group 2 in Figure~\ref{fig:skill_relevance}.
Here, we show the 3 most relevant skills for several ISCO levels of the ``Professionals'' ISCO group.

We notice the following:
First, communication-related skills appear in multiple forms across occupation groups. 
The terms \textit{communication}, \textit{communication sciences}, \textit{communication studies}, \textit{ICT communication protocols}, \textit{manage online communications} and \textit{communication disorders} seem to be closely related. 
Because these skills are defined as distinct skills, each skill receives its own ranking. This concept can appear multiple times. 

Next, ``Nursing professionals'' and ``Nursing and midwifery professionals'' share the same set of relevant skills, which are highly similar to those of their parent group ``Health professionals''. Skills that appear in those groups are the most frequent skills in the parent group.

Finally, the further down the figure we go, the more specialized the skills appear to be, and more specialized skills, such as ``dental studies,'' are more commonly observed in level 4 ISCO groups. A possible explanation for this is that specialized skills do only appear at specialized occupations.

\begin{table*}[]
\resizebox{\textwidth}{!}{\begin{tabular}{@{}ll|l|l@{}}
\toprule
  & Managers & Professionals & Technicians and associate professionals \\ \midrule
1 & Microsoft Office & Network Marketing & Marker Making \\
2 & Service-oriented Modelling & Manage Online Communications & Electronic Communication \\
3 & Communication Principles & Communication & Service-oriented Modelling \\
4 & Electronic Communication & Explain Accounting Records & Education Administration \\
5 & Coordinate Patrols & Accounting & Manage Standard ERP System \\ \midrule
  & Clerical support workers & Service and sales workers & Skilled agricultural, forestry and fishery workers \\ \midrule
1 & Execute Administration & Security Panels & Leadership Principles \\
2 & Perform Clerical Duties & Electronic Communication & Agricultural Information Systems and Databases \\
3 & Microsoft Office & Create Solutions to Problems & Pruning Techniques \\
4 & Education Administration & Execute Administration & Spray Pesticides \\
5 & Human Resource Management & Recreation Activities & Lop Trees \\ \midrule
  & Craft and related trades workers & Plant and machine operators, and assemblers & Elementary occupations \\ \midrule
1 & Attend to Detail in Casting Processes & Mechatronics & Inventory Management Rules  \\
2 & Attention to Detail & Mechanical Engineering & Have Computer Literacy \\
3 & Adobe Illustrator & Electrical Engineering & Carpentry \\
4 & Adobe Photoshop & Operate Soldering Equipment & Place Concrete Forms \\
5 & ML (computer programming) & Act Reliably & Operate on-board Computer Systems \\ \bottomrule
\end{tabular}}
\caption{Five most relevant skills per major ISCO group based in the TF-IDF matric}
\label{tab:skill_relevance}
\end{table*}

\section{Conclusion}\label{chap:conclusion}
In recent years the labor market has changed drastically. 
This is mostly due to increased globalization, a growing working population and disappearing jobs due to digitalization. 
The COVID-19 pandemic has accelerated this change. 
This paper aims to explore algorithmic and data-driven methods for exploring and improving the fit between job seekers and vacancies by modeling skills and occupation data in a knowledge graph. 
Modeling and leveraging relationships between occupations and skills can provide insights for job seekers with existing skill sets. 

After constructing our knowledge graph by relying on the existing ISCO and ESCO taxonomies for occupations and skills, we enrich our KG by relying on job posting data.

We explore our final KG using three different applications.

First, we study link prediction methods for quantifying the relatedness between skills and occupations in Section~\ref{chap:RQ1}. 
We compare and evaluate two different link prediction methods, and find that ``Node2Vec'' performs best. 
Next to quantifying relatedness between occupations and skills for, e.g., ranking skills for an occupation or using as edge weights in our KG, we explore Node2Vec for identifying skills-to-occupation links that are not present in the original KG. 

Next, in Section~\ref{chap:RQ2} we explore our KG for finding efficient job transitions. 
When an individual is searching for a job, knowing which occupations can help the search process. 
In our next application we explore shortest path finding algorithms for identifying potential careerpath prediction.  
We use a skills-based Jaccard similarity metric to model distance between occupations. 
Furthermore, we show examples of job transitions and study properties of our KG by analyzing the distribution of distances between skills and occupations.

Finally, in Section~\ref{chap:RQ3} we study a method to determine which skills are most relevant to different levels of aggregated occupations, using the ISCO taxonomy.
The skill relevance to an ISCO (group) is calculated by taking the frequency of the skills being required for an ISCO (group) with the uniqueness of the skill in the overall ISCO taxonomy. 
Here, the uniqueness is high if a skill occurs more often in one group compared to the other groups. 
The metric that reflects this intuition is called ``TF-IDF.'' 
By doing so we construct a birds-eye view of the labor market. 

The findings from the three sections described above are all variations to the same theme, of finding or enabling the perfect fit between a job seeker and a vacancy, by leveraging skills.

\section{Discussion \& Future research}\label{chap:discussion}
In this paper we present different KG-driven applications for skills-based job matching.
In principle, the methods presented are data-agnostic, as long as similar concepts (occupations and skills) and data (job postings with identified occupations and skills) are available. 
More specifically, we leverage the ISCO and the ESCO taxonomies, which are available in a large number of languages, and are considered standards that are freely available.
Other frameworks could be used as well, where ESCO is widely used in Europe, the O*NET framework~\cite{ONET} is often referred to as the de facto standard in the United States.

The outcome of any research is heavily dependent on the available data. 
In the case of this research this data is preprocessed in a number of steps, one of which is the skill matching step described in Section~\ref{Skill Matching}. 
We opted for a naive character $n$-gram based method for matching surface forms found in a job posting with skill names in ESCO. 
Obviously, more refined methods can be employed, e.g., by considering additional representation of the skill in the job posting (contextual words, occupation), and at the same time additional context at the side of the KG (e.g., skill descriptions, associated occupations, etc.).
In general, this problem of matching can be considered an entity linking task, which is considered out of scope for this application paper. 
Having a flawed knowledge graph as a result of sub-optimal prepossessing does not invalidate the methods used. 
Whichever approach is used to create a knowledge graph, the outcome will never be perfect \cite{10.1145/2623330.2630803}. 

Finally, two out of three applications of our KG are not validated empirically: for both our shortest path finder (Section~\ref{chap:RQ2}) and identifying the most relevant skill per ISCO group (Section~\ref{chap:RQ3}, we focused on the analysis and interpretation of results, omitting a more formal evaluation methodology. For future research it would be interesting to benchmark the current against different career path prediction models.
Validating if, e.g., the discovered paths between occupations indeed are practically the shortest one, requires additional data. 
Unfortunately, no such data was available at the time of writing. 
One place to acquire such data, is, e.g., by collecting data of historic career paths. 
However, collecting such data and composing was determined out of scope for this work. 
The same arose for the method for quantifying the relevance of skills per ISCO group; these aggregated insights were difficult to validate.
We could imagine involving human expert annotators to annotate which skills they deem (most) relevant to a certain ISCO (group). 
However, similarly to the above, collecting and analyzing such data did not fit in the scope of our present work.
In summary, our paper revolves around studying algorithmic methods that aim to help both jobseekers and recruiters find a better match between individuals and occupations, we consider studies with actual end-users out of scope~\cite{10.1145/3298689.3347001}.

\begin{acks}
A special thanks to the thesis supervisors for the project Niels van Weeren and Prof. Aske Plaat as well as everybody at Randstad involved in this project.
\end{acks}

\bibliographystyle{ACM-Reference-Format}
\bibliography{thesis}

\end{document}